\documentstyle[epsfig]{aipproc}
\def\gs{\gtrsim}
\def\ls{\lesssim}

\def\be{\begin{equation}}
\def\en{\end{equation}}                  
\def\bea{\begin{eqnarray}}
\def\ena{\end{eqnarray}}
\newcommand{\bi}[1]{\mbox{\boldmath$#1$}}
\newcommand{\av}[1]{\langle{#1}\rangle}

\begin{document}
\title{Dynamics of Highly Supercooled Liquids}

\author{R. Yamamoto and A. Onuki}
\address{Department of Physics, Kyoto University, Kyoto 606-8502, Japan}

\maketitle

\begin{abstract}

The diffusivity of tagged particles is demonstrated 
to be heterogeneous on time scales comparable to or less than 
the structural relaxation time
in a highly supercooled liquid via 3D molecular dynamics simulation. 
The particle motions in the relatively active regions 
dominantly contribute to the mean square displacement,  
giving rise to a diffusion constant systematically larger than 
the Stokes-Einstein value.
The van Hove self-correlation function $G_s(r,t)$ is shown 
to have a large $r$ tail which can be scaled in terms of 
$r/t^{1/2}$ for $t \ls 3 \tau_\alpha$, where
$\tau_\alpha \cong$ the stress relaxation time. 
Its presence indicates heterogeneous diffusion in the active regions.
However, the diffusion process eventually becomes homogeneous 
on time scales longer than the life time of the heterogeneity 
structure ($\sim 3 \tau_{\alpha}$).

\end{abstract}

\section*{Introduction}

Molecular dynamics (MD) simulations can be powerful tools to
gain insights into relevant physical processes in highly
supercooled liquids.
In particular, a number of recent MD simulations have 
detected dynamic heterogeneities in supercooled model 
binary mixtures \cite{Muranaka,Harrowell,Yamamoto_Onuki1,Yamamoto_Onuki98,yo_prl,Donati}. 
That is, rearrangements of particle configurations in 
glassy states are cooperative, involving many molecules,  
owing to configuration restrictions.
Recently, we succeeded in quantitatively characterizing 
the dynamic heterogeneities in 
two (2D) and three dimensional (3D) model fluids
via MD simulations.
We examined bond breakage processes among 
adjacent particle pairs and found that the broken bonds in an  
appropriate time interval ($\sim$ the stress relaxation time
or the structural $\alpha$ relaxation time $\tau_{\alpha}$) 
are very analogous to the critical fluctuations 
in Ising spin systems with their structure factor being 
excellently fitted to the Ornstein-Zernike form 
\cite{Yamamoto_Onuki1,Yamamoto_Onuki98}.
The correlation length $\xi$ thus obtained  
is related to $\tau_{\alpha}$ via the dynamic scaling law,  
$\tau_{\alpha} \sim \xi^z$,  with 
$z=4$ in 2D and $z=2$ in 3D. 
The heterogeneity structure in the bond breakage is essentially 
the same as that in jump motions of particles from cages 
or that in the local diffusivity, as will be discussed below. 
In this paper, we investigate heterogeneities of tagged 
particle motions in a 3D supercooled liquid \cite{yo_prl}.

In a wide range of liquid states, the Stokes-Einstein relation 
$D\eta a/ k_BT=const.$  has been  successfully 
applied  between the translational diffusion constant $D$ 
of a tagged particle and the viscosity $\eta$ even 
when the tagged particle diameter $a$ 
is of the same order as that of solvent molecules. 
However, this relation is systematically violated 
in fragile supercooled liquids 
\cite{Yamamoto_Onuki98,yo_prl,Ediger,Sillescu,Ci95,Mountain,Perera_PRL98}.
The diffusion process in supercooled liquids is thus 
not well understood. 
In particular, Sillescu {\it et al.} 
observed the power law behavior 
$D \propto \eta^{-\nu}$ with $\nu \cong 0.75$ 
at low temperatures \cite{Sillescu}. 
Furthermore, Ediger {\it et al.} found that 
smaller probe particles exhibit a more pronounced increase of  
$D\eta/T \propto D/D_{SE}$ with decreasing $T$ \cite{Ci95}, 
where $D_{SE} \sim  k_BT/2\pi \eta a$ is the Stokes-Einstein  
diffusion constant. 
In such experiments the viscosity changes over 12 decades with 
decreasing  $T$, while the ratio $D/D_{SE}$ increases from 
order 1 up to order $10^2 \sim 10^3$.  
The same tendency has been detected by molecular dynamics simulations 
in a 3D binary mixture with $N=500$ 
particles \cite{Mountain} and in a 2D
binary mixture with $N=1024$ \cite{Perera_PRL98}. 
In our recent 3D simulation with $N= 10^4$ \cite{Yamamoto_Onuki98,yo_prl}, 
$\eta$ and $D$ both varied over 4 decades and 
the power law behavior $D \propto \eta^{-0.75}$ has been observed. 
Many authors have attributed the origin of the breakdown to 
heterogeneous coexistence of relatively active and inactive regions, 
among which the local diffusion constant is expected 
to vary significantly \cite{Sillescu,Ci95,St94,Tarjus,Oppen}.
The aim of this paper is to demonstrate via MD simulation that 
the diffusivity of the particles is indeed heterogeneous 
on time scales shorter than $\tau_{\alpha}$
but becomes homogeneous on time scales 
much longer than $\tau_{\alpha}$.

\section*{Simulation method and results}

Our 3D binary mixture is composed of two 
atomic species, $1$ and $2$, with $N_{1}=N_{2}=5000$ particles
with the system linear dimension $L=V^{1/3}$ being fixed at 
$23.2\sigma_1$  \cite{Bernu}.  
They interact via the soft-core potentials  
$ v_{ab}(r)= \epsilon (\sigma_{ab}/r)^{12}$ with 
$\sigma_{ab}=(\sigma_{a}+\sigma_{b})/2$,   
where $r$ is the distance between two particles and 
$a,b \in \{1,2\}$.  
The interaction is truncated at $r =3\sigma_{1}$.
The mass ratio is  $m_{2}/m_{1}=2$. 
The size ratio is  $\sigma_{2}/\sigma_{1}=1.2$, 
which is known to prevent crystallization \cite{Muranaka,Bernu}.  
No tendency of phase separation is 
detected at least in our computation times. 
We fix the particle density at a very high value of 
$(N_1+N_2)/V =0.8/\sigma_{1}^{3}$, 
so the particle configurations are severely restricted or jammed. 
We will measure space and time in units of 
$\sigma_1$ and  $\tau_0=({m_{1}\sigma_{1}^{2}/\epsilon})^{1/2}$. 
The temperature $T$ will be measured  in units of $\epsilon/k_B$,  
and the viscosity $\eta$ in units of $\epsilon \tau_0/\sigma_1^3$. 
The time step $\Delta t = 0.005$ is used.
In our systems the structural relaxation time
becomes very long at low temperatures.
Therefore, very long annealing times ($2.5\times10^5$ for $T=0.267$) 
are chosen in our case. 
For $T\ge0.267$, no appreciable aging 
(slow equilibration) 
effect is detected 
in various quantities such as the pressure or the density time 
correlation function, whereas at $T=0.234$, a small aging effect 
remains in the density time 
correlation function.

\begin{figure}[t]
\centerline{\epsfig{file=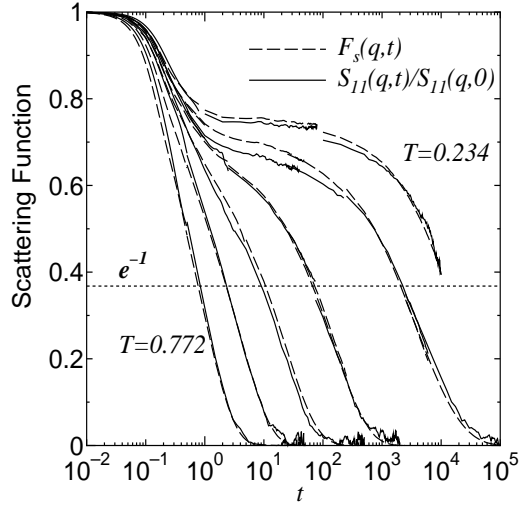,height=2.7in,width=2.7in}}
\caption{\protect
Coherent and incoherent intermediate scattering functions 
for various temperatures with $q=2\pi$ (a peak wavenumber in $S_{11}(q,0)$).
$T=0.234$, $0.267$, $0.306$, $0.352$, $0.473$, and $0.772$ from right.
}
\label{taua_vis}
\end{figure}

Let us first consider the incoherent density correlation function, 
$F_s(q,t)= \av{\sum_{j=1}^{N_1} 
\exp {[i{\bf q}\cdot \Delta{\bi r}_j(t)]}}/N_1$ 
for the particle species 1, where 
$\Delta{\bi r}_j(t)={\bi r}_j(t)- {\bi r}_j(0)$ 
is the displacement vector of the $j$-th particle. 
The $\alpha$ relaxation time $\tau_{\alpha}$ is then defined by 
$F_s(q,\tau_{\alpha})= e^{-1}$ at $q = 2\pi$ for various $T$. 
We also calculate the coherent time correlation function,  
$S_{11}(q,t)= 
\langle 
n_1({\bi q}, t)n_1({-{\bi q}}, 0)
\rangle$, where 
$n_1({\bi q}, t)=\sum_{j=1}^{N_1} 
\exp {[i{\bf q}\cdot {\bi r}_j(t)]}$ is the Fourier component 
of the density fluctuations of the particle species 1.  
The decay profiles of $S_{11}(q,t)$ at its first peak 
wave number $q=q_m \sim 2\pi$ and $F_s(q,t)$ at $q=2\pi$ 
nearly coincide in the whole time region studied 
($t < 2 \times 10^5$) within $5\%$ as shown in Fig. 1.
Hence $S_{11}(q_m,\tau_{\alpha})/S_{11}(q_m,0) \cong e^{-1}$ 
holds for any $T$ in our simulation. 
Such agreement is not obtained for other wave numbers, however.
These results are  consistent with those 
for a Lennard-Jones binary mixture \cite{Kob2}. 
Furthermore, some neutron-spin-echo experiments \cite{Mezei} 
showed that the decay time of $S_{11}(q_m,t)$ is nearly 
equal to the stress relaxation time and as a result 
the viscosity $\eta$ is of order $\tau_{\alpha}$. 
In agreement with this experimental result, 
we obtain a simple linear relation in our simulations \cite{yo_prl}, 
\be
\tau_{\alpha}\cong  (A_\eta /q_m^2)\eta /T. 
\label{eq:1}
\en
The coefficient $A_{\eta}$ is close to $2\pi$ in our system. 
Here, we may define a $q$-dependent 
relaxation time $\tau_q$ by $F(q,\tau_q)=e^{-1}$. 
Thus, particularly at the peak wave number $q=q_m$, 
the effective diffusion constant defined by 
$D_q \equiv 1/q^2\tau_q$ 
is given by the Stokes-Einstein form even in 
highly supercooled liquids. 
However, notice that the usual diffusion constant  
is the long wavelength limit,  
$D= \lim_{q\rightarrow 0} D_q$. 
It is usually calculated from the mean square  
displacement, 
$
\av{(\Delta{\bi r}(t))^2}= \av{
\sum_{j=1}^{N_1}(\Delta{\bi r}_j(t))^2}/N_1.
$
The crossover of this quantity from the plateau behavior 
arising from motions in transient cages to the diffusion 
behavior $6Dt$ has been found to take place around 
$t \sim  0.1 \tau_{\alpha}$ \cite{Yamamoto_Onuki98}. 
In Fig.2, we plot $D\tau_{\alpha}$ versus $\tau_{\alpha}$, 
which clearly indicates breakdown of the Stokes-Einstein 
relation in agreement with the experimental trend.

\begin{figure}[t]
\centerline{\epsfig{file=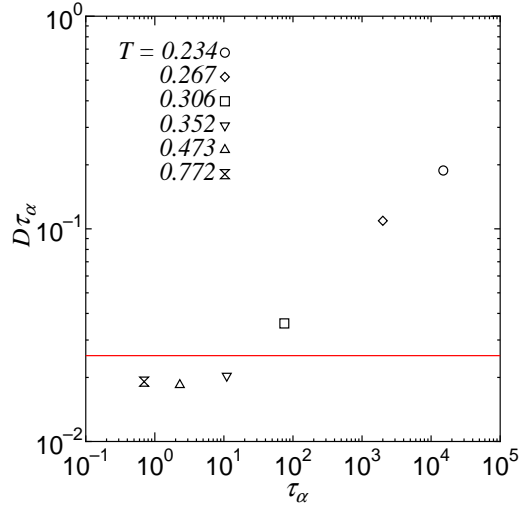,height=2.7in,width=2.7in}}
\caption{\protect
$D\tau_\alpha$ versus $\tau_\alpha$ in a quiescent supercooled liquid.
The solid line represents the Stokes-Einstein value 
$D_{SE}\tau_{\alpha}=(2\pi)^{-2}$ arising from Eq.(1).}
\label{taua_dtaua}
\end{figure}

\begin{figure}[t]
\centerline{\epsfig{file=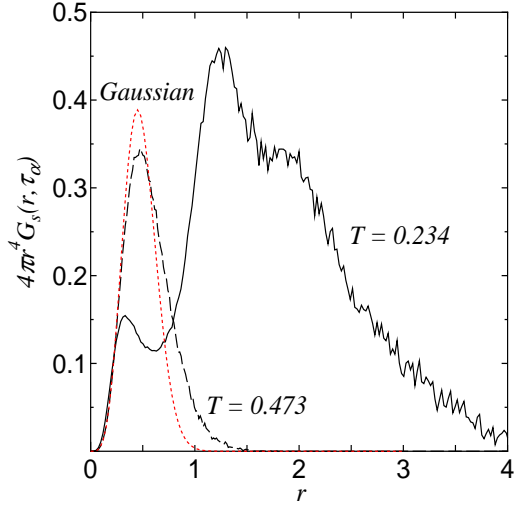,height=2.7in,width=2.7in}}
\caption{\protect
$4\pi r^4 G_s(r,t)$ versus $r$ at $t=\tau_{\alpha}$.
The solid line is for $T=0.267$ and the broken line is 
for $T=0.473$. 
The dotted line represents the Brownian motion result.
The peaks at $r \simeq 1.2$ and $2$ in the solid line 
arise from hopping processes in our system at $T=0.267$.
Note that the areas below the curves give $6D\tau_{\alpha}$.
}
\label{6dta}
\end{figure}

\begin{figure}[t]
\centerline{
\epsfig{file=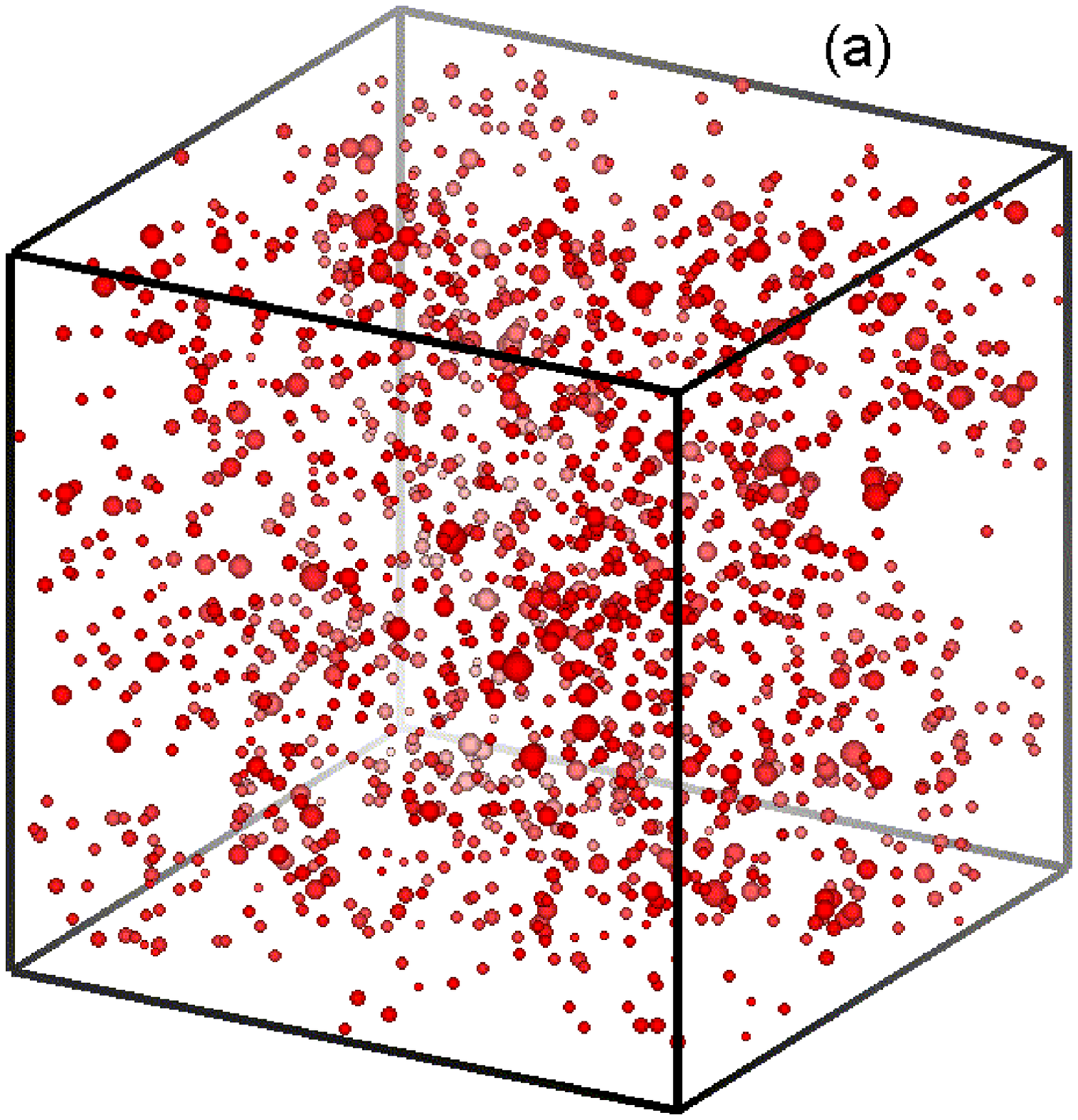,height=1.9in,width=1.9in}
\epsfig{file=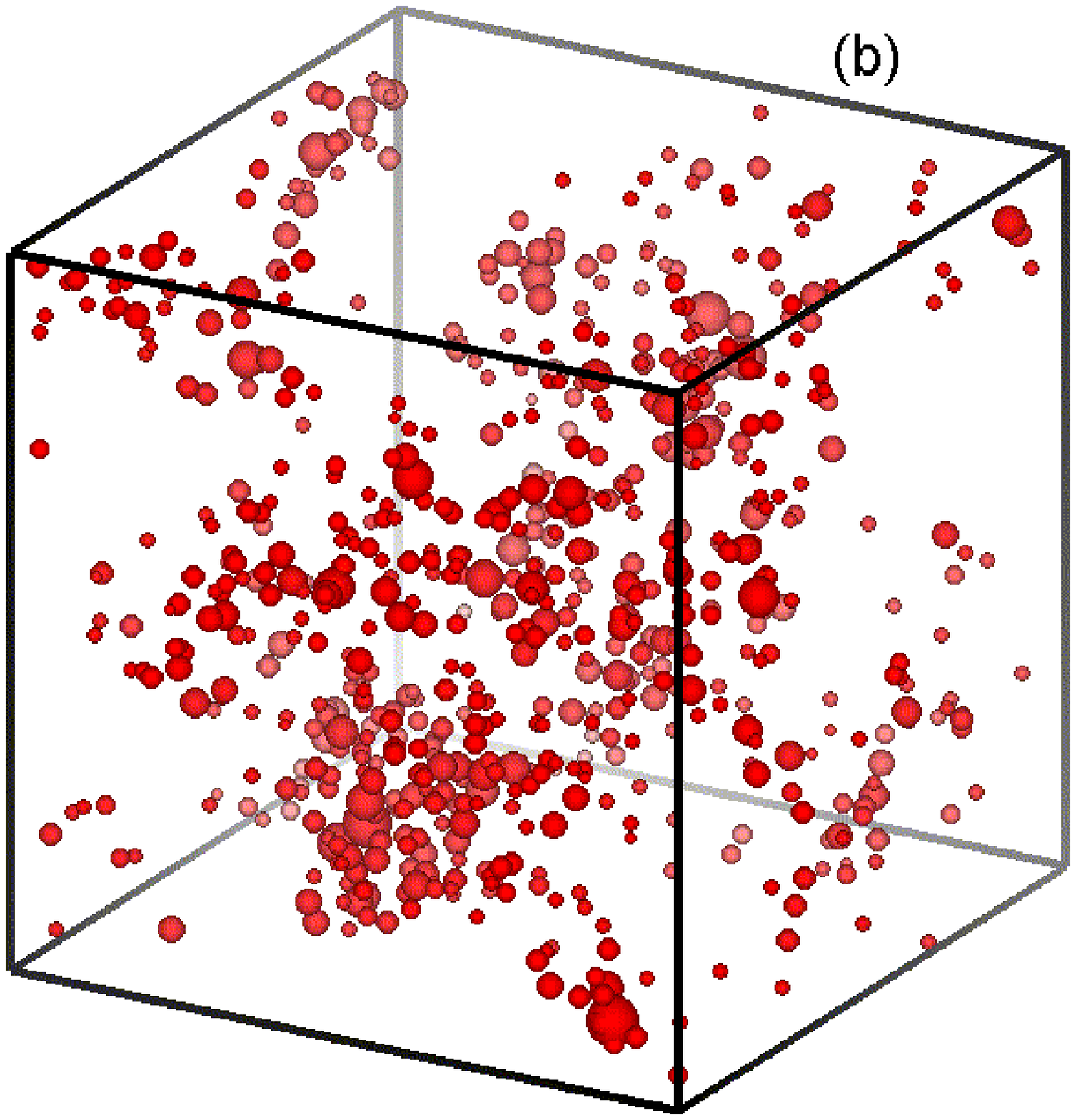,height=1.9in,width=1.9in}
\epsfig{file= 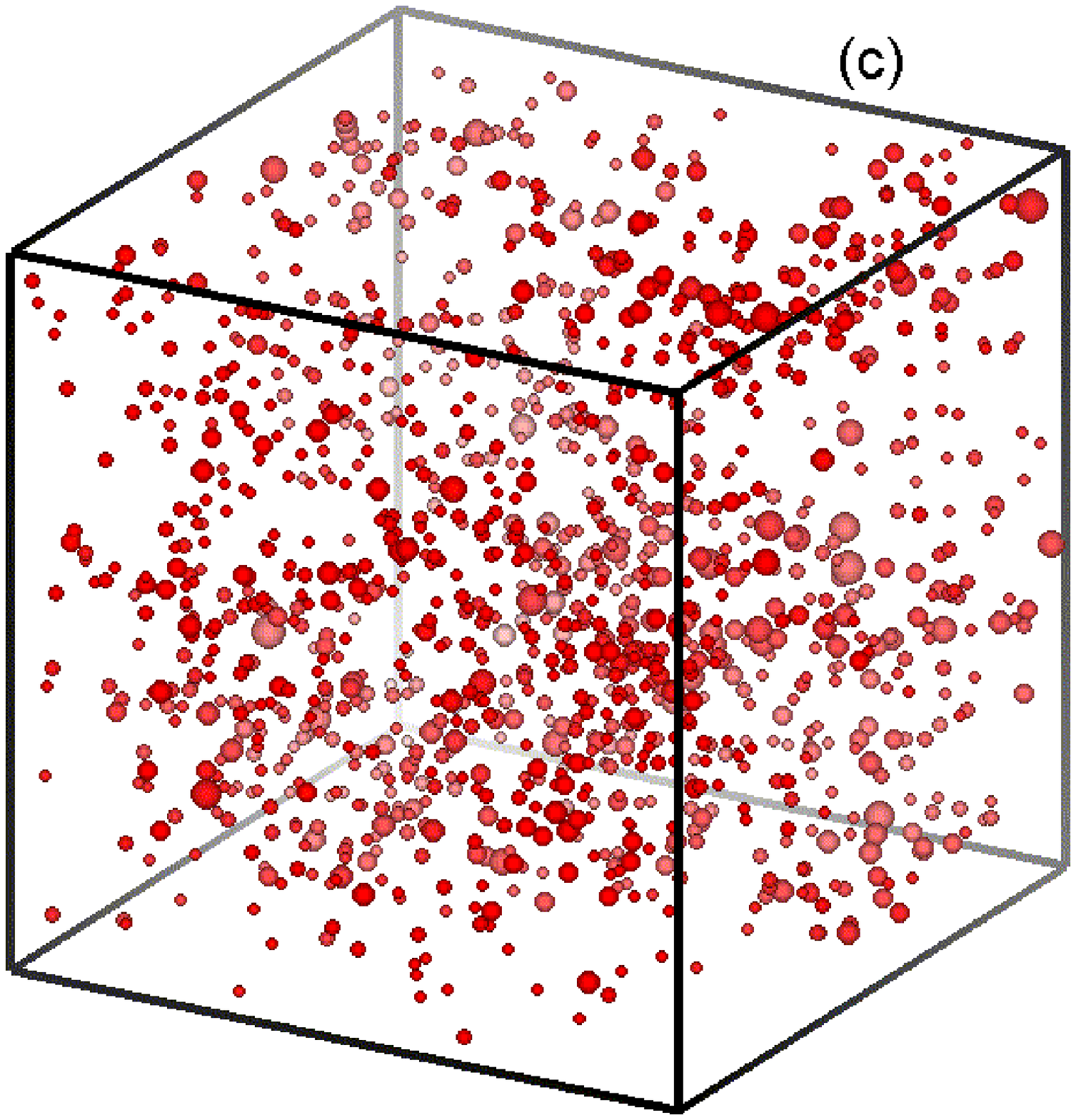,height=1.9in,width=1.9in}
}
\caption{
Mobile particles of the species 1 with a time interval 
$t=\tau_{\alpha}$ at $T=0.473$ (a) and $T=0.267$ (b) 
and with $t=10\tau_{\alpha}$ at $T=0.267$ (c). 
The radii of the spheres are 
$|\Delta{\bi r}_j(t)|/\sqrt{\av{(\Delta{\bi r}(t))^2}}$ 
and the centers are at 
$\frac{1}{2} [ {\bi r}_j(t_0)+{\bi r}_j(t_0+t)]$.  
The system linear dimension is $L=23.2$.
The darkness of the spheres represents the depth in the 3D space.
}
\label{space}
\end{figure}

To examine the diffusion process in more detail, 
we consider the van Hove self-correlation function,  
$
G_{s}(r,t)=   
\av{ \sum_{j=1}^{N_1}\delta (\Delta{\bi r}_j(t)-{\bi r}) }/N_1 .
$  
Then,  
\be
F_s(q,t) = 
\int_0^\infty dr \frac{\sin(qr)}{qr}4\pi r^2 G_s(r,t)
\label{eq:2}
\en
is the 3D Fourier transformation of $G_{s}(r,t)$. 
At the peak 
wavenumber $q=2\pi$, the integrand in Eq.(2) vanishes 
at $r=1$, and the integral in the region $r<1$ is confirmed 
to dominantly determine the decay of $F_s(2\pi,t)$. 
On the other hand, the mean square displacement 
\be
\av{(\Delta{\bi r}(t))^2}= 
\int_0^\infty dr 4\pi r^4 G_s(r,t)
\label{eq:3}
\en  
is determined by the particle motions out of the cages for 
$t \gs \tau_{\alpha}$ in glassy states. 
In Fig.3, we display $4\pi r^4 G_s(r,\tau_{\alpha})$ versus $r$, 
where $\tau_{\alpha}=3.2$ and $2000$ for $T=0.473$ and $0.267$, 
respectively. 
These curves may be compared with the Gaussian (Brownian motion) 
result, $(2/\pi)^{1/2}\ell^{-3}r^4 \exp (-r^2/2\ell^2)$,  
where $3 \ell^2= 6D_{SE}\tau_{\alpha} = 3/2\pi^{2}$ is the 
Stokes-Einstein mean square displacement. 
Because the areas below the curves 
give $6D\tau_{\alpha}$, we recognize that the particle motions 
over large distances $r > 1$ are much enhanced at low $T$, 
leading to the violation of the Stokes-Einstein relation.

We then visualize the heterogeneity of the diffusivity.
To this end, we pick up mobile particles of the species 1 
with $|\Delta{\bi r}_j(t))| > \ell_c$ in a time interval 
$[t_0, t_0+t]$ and number them as $j=1, \cdots, N_m$. 
Here $\ell_c$ is defined such that the sum of 
$\Delta{\bi r}_j(t)^2$ of the mobile particles is 
$66\%$ of the total sum ($\cong 6DtN_1$ for $t \gs 0.1\tau_{\alpha}$).
In Fig.4, these particles are drawn as spheres with radii 
\be
a_j(t) \equiv |\Delta{\bi r}_j(t)|/\sqrt{\av{(\Delta{\bi r}(t))^2}}
\label{eq:4}
\en 
located at ${\bi R}_j(t)\equiv  
\frac{1}{2} [ {\bi r}_j(t_0)+{\bi r}_j(t_0+t)]$
in time intervals $[t_0, t_0+\tau_{\alpha}]$ 
for $T=0.473$ (a) and $0.267$ (b)
and in $[t_0, t_0+10\tau_{\alpha}]$ for $T=0.267$ (c).
The the mobile particle number $N_m$ is
$1595$ in (a), $725$ in (b), and $1316$ in (c), respectively.  
Here the Gaussian results is $N_m=1800$. 
The ratio of the second moments 
$c_2 \equiv \sum_{j=1}^{N_m} a_j(t)^2/\sum_{j=1}^{N_1} a_j(t)^2$ 
is held fixed at $0.66$, while the ratio of the fourth moments 
$c_4 \equiv \sum_{j=1}^{N_m} a_j(t)^4/\sum_{j=1}^{N_1} a_j(t)^4 $ 
turns out to be close to 1 as 
$c_4=0.89$ in (a), $0.92$ in (b), and $0.90$ in (c). 
The mobile particles are homogeneously distributed for $T=0.473$
at $\tau_{\alpha}$, whereas for $T=0.267$, the heterogeneity 
is significant at $\tau_{\alpha}$, 
but is much decreased at $10 \tau_{\alpha}$.
In fact, the variance defined by 
$
{\cal V} \equiv 
N_m \sum_{j=1}^{N_m} a_j(t)^4/(\sum_{j=1}^{N_m} a_j(t)^2)^2  -1 
$
is $0.27$ in (a), $0.41$ in (b), and $0.32$ in (c). 
Note that the statistical average of 
$\cal V$ (taken over many initial times $t_0$) is related to  
the non-Gaussian parameter 
$A_2\equiv 3\av{\Delta{\bi r}(t)^4}/5\av{\Delta{\bi r}(t)^2}^2 - 1 
= 3N_1\av{\sum_{j=1}^{N_1} a_j(t)^4}/5(\av{\sum_{j=1}^{N_1} a_j(t)^2})^2-1$
by 
\be
\av{{\cal V}} \cong (5 \av{c_4}\av{N_m}/3c_2^2 N_1) (1+ A_2(t)) - 1,
\en
where the deviations $c_4-\av{c_4}$ and $N_m/\av{N_m}-1$ 
are confirmed to be very small for large $N_1$ 
and  are thus neglected.  
We may also conclude that the significant rise of $A_2(t)$ 
in glassy states originates from the heterogeneity in accord with 
some experimental interpretations \cite{Zorn}.

We next consider the Fourier component of 
the {\it diffusivity} density defined by  
\be
{\cal D}_{{\bi q}}(t_0, t) \equiv 
\sum_{j=1}^{N_m} a_j(t)^2 \exp [-i{\bi q}\cdot {\bi R}_j(t)],
\label{eq:5}
\en
which depends on the initial time $t_0$ and the final time 
$t_0+t$. 
The correlation function 
$S_{{\cal D}}(q,t, \tau) = 
\av{{\cal D}_{{\bi q}}(t_0 +\tau, t){\cal D}_{-{\bi q}}(t_0 , t)}$ 
is then obtained after averaging over many initial states.
We plot $S_{{\cal D}}(q,\tau_{\alpha},0)$ in Fig. 5 (a).
The heterogeneity structure $S_{b}(q,\tau_{\alpha},0)$ 
of the bond breakage \cite{Yamamoto_Onuki1,Yamamoto_Onuki98} with a time interval 
of $\tau_{\alpha}$ is also plotted in Fig. 5 (b).
It is confirmed that $S_{\cal D}(q,\tau_{\alpha},0)$ tends to 
its long wavelength limit for $q \ls \xi^{-1}$, 
where $\xi$ coincides with the correlation length obtained from 
$S_b(q,\tau_{\alpha},0)$.
As the difference $\tau$ of the initial times increases 
with fixed $t=\tau_{\alpha}$, 
$S_{{\cal D}}(q, \tau_{\alpha}, \tau)$ relaxes as 
$\exp [ - (\tau / \tau_h )^c ]$ for $q \ls \xi^{-1}$, 
where $c \sim 0.5$ at $T=0.267$ and 
$\tau_h \sim 3 \tau_{\alpha}$ is the life time of the 
heterogeneity structure. 
The two-time correlation function among the broken bond density 
\cite{Yamamoto_Onuki1,Yamamoto_Onuki98} 
also relaxes with $\tau_h$ in the same manner.

\begin{figure}[t]
\centerline{
\epsfig{file=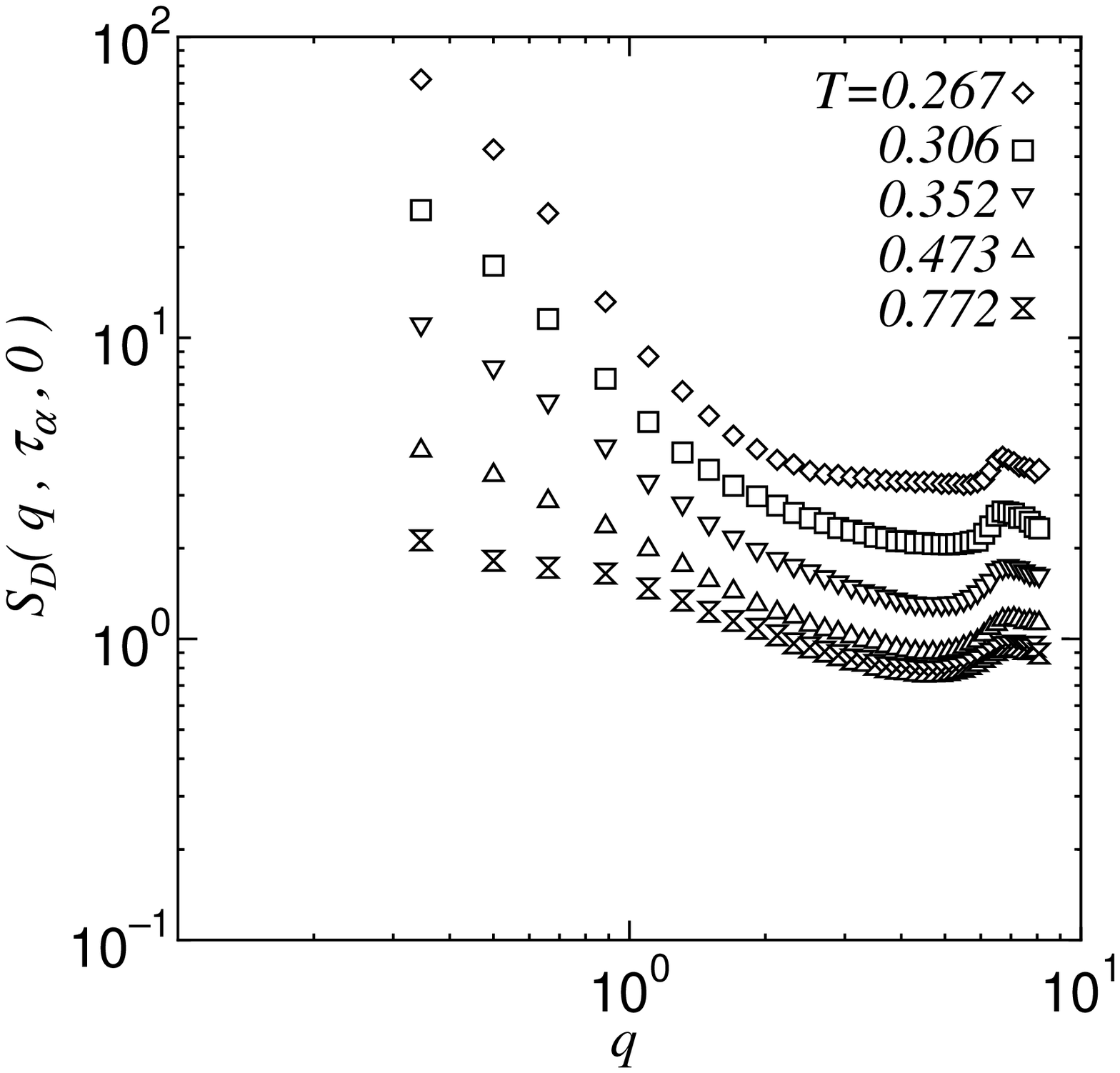,height=2.7in,width=2.7in}
\epsfig{file=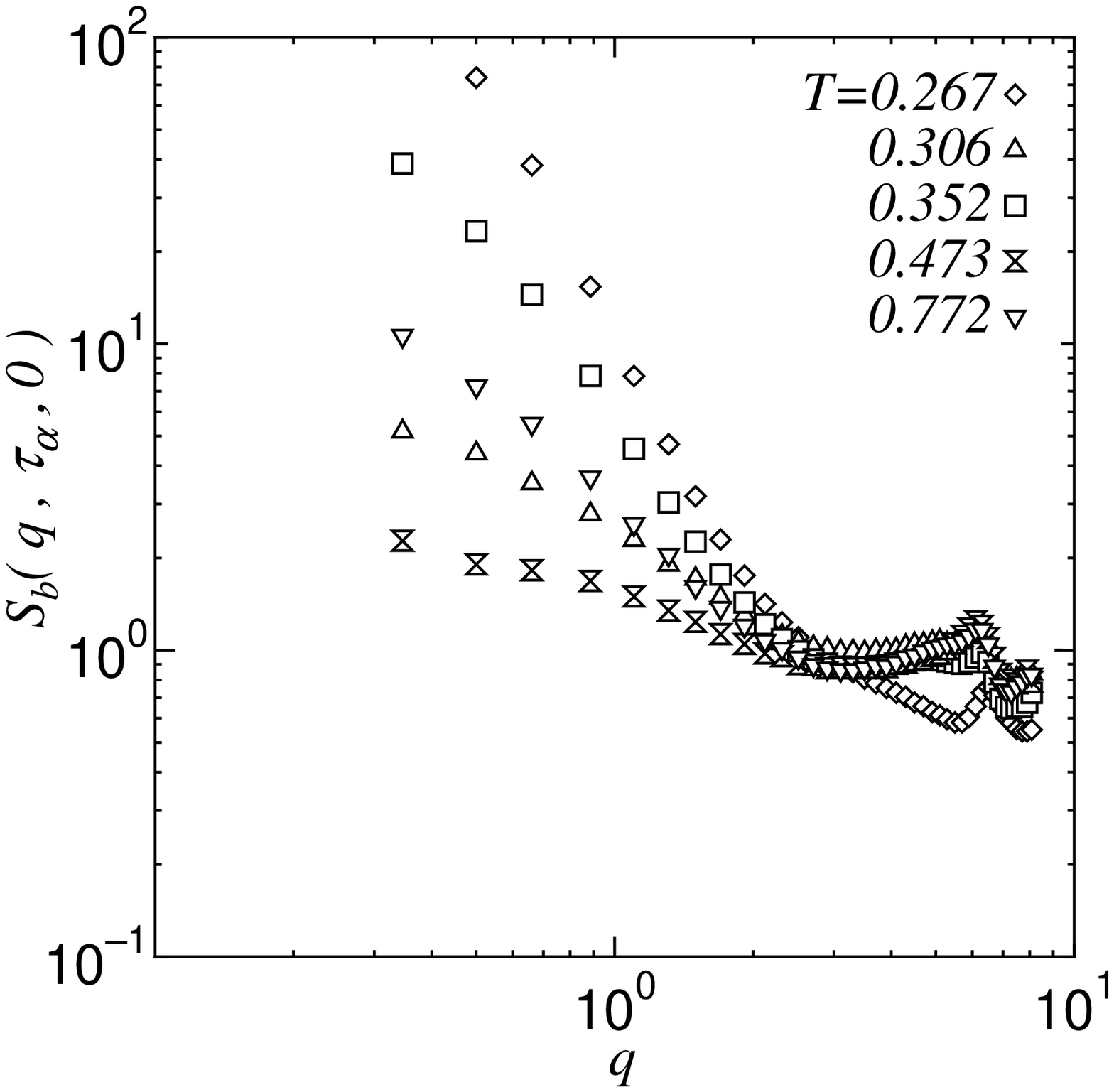,height=2.7in,width=2.7in}
}
\caption{
The correlation functions 
$S_{{\cal D}}(q,\tau_{\alpha},0)$ (a) 
and $S_{b}(q,\tau_{\alpha},0)$ (b) with $q=2\pi$.
}
\label{sk}
\end{figure}

We naturally expect that the distribution of the 
particle displacement 
$\Delta{\bi r}_j(t)$ in the active regions
should be characterized by the local 
diffusion constant $D({\bi x},t)$ dependent on the spatial 
position ${\bi x}=(x,y,z)$ and the time interval $t$. 
The van Hove correlation function $G_s(r,t)$ may then be 
expressed as the spatial average of a local function $G_s({\bi x},r,t)$, 
which is given by 
$[4\pi D({\bi x},t)t]^{-3/2}\exp [ -{r^2}/4D({\bi x},t)t ]$. 
To check this conjecture, we plot the scaled function 
$\sqrt{6Dt} 4\pi r^2 G_s(r,t)$ versus $r^*= r/\sqrt{6Dt}$ in Fig. 6. 
The areas below the curves are fixed at 1. 
At relatively short times $t \ls 3 \tau_{\alpha}$, the curves 
in the region $r \gs 1$ or $r^* \gs (6Dt)^{-1/2}$,
which give dominant contributions to $\av{(\Delta{\bi r}(t))^2}$,
tend to a master curve quite different from the rapidly decaying 
Gaussian tail.  
Note that in each curve the position of the peak at larger $r^*$ 
corresponds to $r \cong 1$. 
This asymptotic law is consistent with the picture of the 
space-dependent diffusion constant in the active regions. 
It is also important that the heterogeneity structure 
remains unchanged in the time region $t \ls \tau_h \sim 3\tau_{\alpha}$.   
At longer times $t \gs 10\tau_{\alpha}$, the curves approach the 
Gaussian form as can be seen in the inset of Fig.5.
Of course, $4\pi r^2G_s(r,t)$ for $r<1$ does not scale 
in the above manner, because it is the probability density 
of a tagged particle staying  within a cage.  
This short-range behavior determines the decay of $F_s(2\pi,t)$ 
as noted below Eq.(2).

\begin{figure}[t]
\centerline{\epsfig{file=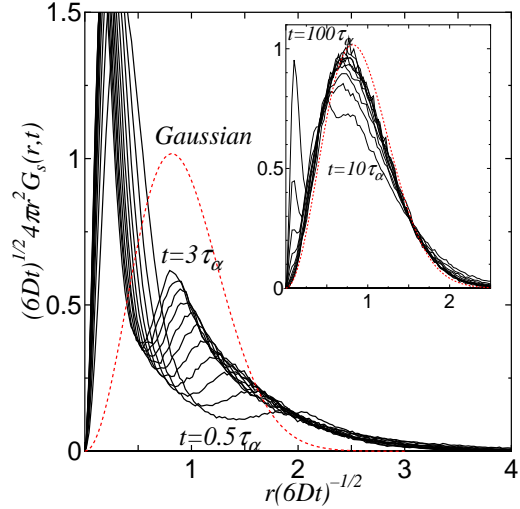,height=2.7in,width=2.7in}}
\caption{\protect
A test of the scaling plot 
$\sqrt{6Dt} 4\pi r^2 G_s(r,t)$ versus $r/\sqrt{6Dt}$ 
for $0.5 \tau_{\alpha} < t < 3\tau_{\alpha}$ 
($t=(0.5+0.25n)\tau_{\alpha}$, $n=0,1,2,\cdots,10$).
The inset shows the curves at longer times, 
$10\tau_{\alpha} <t<100\tau_{\alpha}$ 
($t=10n\tau_{\alpha}$, $n=1,2,\cdots,10$), 
where the heterogeneity effect is smoothed out. 
The dotted lines are the Gaussian form. }
\label{scale}
\end{figure}

\section*{Concluding remarks}

In our previous studies \cite {Yamamoto_Onuki1,Yamamoto_Onuki98}, 
we performed extensive MD simulations and 
identified {\it weakly bonded} or {\it relatively active} 
regions from breakage of appropriately defined bonds.
We also found that the spatial distributions of such regions resemble 
the critical fluctuations in Ising spin systems,
so the correlation length $\xi$ can be determined.
It grows up to the system size as $T$ is lowered, but no
divergence seems to exist at nonzero temperatures.
In the present work, we have demonstrated that 
the diffusivity in supercooled 
liquids is spatially heterogeneous on time scales shorter than 
$3\tau_{\alpha}$, which leads to the breakdown of the 
Stokes-Einstein relation \cite{yo_prl}.
The heterogeneity detected is essentially the same as 
that of the bond breakage in our previous works 
\cite{Yamamoto_Onuki1,Yamamoto_Onuki98}. 
We should then investigate how the heterogeneity 
arises and influences observable quantities in 
more realistic glass-forming fluids 
with complex structures.

\section*{ACKNOWLEDGMENTS}

We thank  Prof. T. Kanaya and Prof. M.D. Ediger 
for helpful discussions. 
This work is supported by Grants in Aid for Scientific 
Research from the Ministry of Education, Science and Culture.


\begin{references}


\bibitem{Muranaka} Muranaka, T.,  and Hiwatari, Y.,
{\it Phys. Rev. E} {\bf 51}, R2735-R2738 (1995); 
Muranaka, T. and Hiwatari, Y.,
{\it J. Phys. Soc. Jpn.} {\bf 67}, 1982-1987 (1998).

\bibitem{Harrowell} Hurley, M.M., and Harrowell, P., 
{\it Phys. Rev. E} {\bf 52}, 1694-1698 (1995); 
Perera, D.N., and Harrowell, P.,
{\it Phys. Rev. E}  {\bf 54}, 1652-1662 (1996). 

\bibitem{Yamamoto_Onuki1} Yamamoto, R., and Onuki, A.,
{\it J. Phys. Soc. Jpn.} {\bf 66}, 2545-2548 (1997);
{\it Europhys. Lett.} {\bf 40}, 61-66 (1997);
Onuki, A., and Yamamoto, R.,
{\it J. Non-Cryst. Solids} {\bf 235 -237}, 34-40 (1998).

\bibitem{Yamamoto_Onuki98} 
Yamamoto, R., and Onuki, A.
{\it Phys. Rev. E} {\bf 58}, 3515-3529 (1998).

\bibitem{yo_prl} Yamamoto, R., and Onuki, A.,
{\it Phys. Rev. Lett.} in press (1998).

\bibitem{Donati} Kob, W., {\it et al.},
{\it Phys. Rev. Lett.} {\bf 79}, 2827-2830 (1997); 
Donati, C., {\it et al.},
{\it Phys. Rev. Lett.} {\bf 80}, 2338-2341 (1998).



\bibitem{Ediger}  Ediger, M.D., Angell, C.A., and Nagel, S.R.,
{\it J. Phys. Chem.} {\bf 100}, 13200-13212 (1996).

\bibitem{Sillescu}  Fujara, F., Geil, B., Sillescu, H., and Fleischer, G.,
{\it Z. Phys. B}  {\bf 88}, 195-204 (1992);  
Chang, I., {\it et al.},
{\it  J. Non-Cryst. Solids} {\bf 172-174}, 248-255 (1994).

\bibitem{Ci95} Cicerone, M.T., Blackburn, F.R., and Ediger, M.D.,
{\it Macromolecules} {\bf 28}, 8224-8232 (1995); 
Cicerone, M.T., and Ediger, M.D.,
{\it J. Chem. Phys.} {\bf 104}, 7210-7218 (1996).

\bibitem{Mountain} Thirumalai, D., and Mountain, R.D.,
{\it  Phys. Rev. E} {\bf 47}, 479-489 (1993).

\bibitem{Perera_PRL98} Perera, D., and Harrowell, P.,
{\it Phys. Rev. Lett.} {\bf 81}, 120-123 (1998).

\bibitem{St94} Stillinger, F.H., and Hodgdon, A.,
{\it Phys. Rev. E} {\bf 50}, 2064-2068 (1994).

\bibitem{Tarjus} Tarjus, G., and Kivelson, D.,
{\it J. Chem. Phys.}  {\bf 103}, 3071-3073 (1995).  

\bibitem{Oppen}  Liu, C.Z. -W., and Oppenheim, I.,
{\it Phys. Rev. E} {\bf 53}, 799-802 (1996). 
  
\bibitem{Bernu} Bernu, B., Hiwatari, Y., and Hansen, J.P.,
{\it J. Phys. C} {\bf 18}, L371-L376 (1985); 
Bernu, B., Hansen, J.P., Hiwatari, Y., and Pastore, G.,
{\it Phys. Rev. A}  {\bf 36}, 4891-4903 (1987); 
Matsui, J., Odagaki, T., and Hiwatari, Y.,
{\it Phys. Rev. Lett.} {\bf 73}, 2452-2455 (1994). 

\bibitem{Kob2}  Kob, W., and Andersen, H.C.,
{\it Phys. Rev. E} {\bf 52}, 4134-4153 (1995).

\bibitem{Mezei}  Mezei, F., Knaak, W., and  Farago, B.,
{\it Phys. Rev. Lett.} {\bf 58},  571-574 (1987); 
Richter, D., Frick, R., and  Farago, B.,
{\it Phys. Rev. Lett.} {\bf 61},  2465-2468 (1988).

\bibitem{Zorn} Zorn, R.,
{\it Phys. Rev. B}  {\bf 55}, 6249-6259 (1987);
Kanaya, T., Tsukushi, I., and Kaji, K., 
{\it Prog. Theor. Phys. Supplement} {\bf 126}, 133-140 (1997).

\end{references}
\end{document}